\documentclass[10pt,conference]{IEEEtran}

\usepackage[utf8]{inputenc}
\usepackage[T1]{fontenc}
\usepackage{amsmath,amssymb}
\usepackage{booktabs}
\usepackage{array}
\usepackage{tabularx}
\usepackage{enumitem}
\usepackage[hidelinks]{hyperref}
\hypersetup{pdftitle={Can LLM Agents Learn to Test? Maintaining Scoped Experience as Testing Evidence for Mobile GUI Testing},pdfauthor={Zimu Liu, Xiaoxi Guo, Xiaolu Tang, Liang Yu},pdfsubject={Mobile GUI testing: experience maintenance design and case study}}
\usepackage{tikz}
\usetikzlibrary{automata,positioning,arrows.meta,fit}

\newcolumntype{L}[1]{>{\raggedright\arraybackslash}p{#1}}
\newcolumntype{Y}{>{\raggedright\arraybackslash}X}

\newcommand{\tool}{\textsc{AdaptiTest}}

\begin{document}

\title{Can LLM Agents Learn to Test? Maintaining Scoped Experience as Testing Evidence for Mobile GUI Testing}
\author{\IEEEauthorblockN{Zimu Liu, Xiaoxi Guo, Xiaolu Tang, Liang Yu}
\IEEEauthorblockA{Ant Group\\
\footnotesize \{liuzimu.lzm, xiaoxi.gxx, lance.txl, yuriy.yl\}@antgroup.com}}
\maketitle
\bstctlcite{IEEEexample:BSTcontrol}
\thispagestyle{plain}
\pagestyle{plain}

\begin{abstract}
Repeated mobile GUI tests can reuse routes, locators, and recovery strategies from earlier executions. However, a route that worked in one workflow or app version may be irrelevant or misleading in another. We study how an LLM-based tester can retain useful experience while withholding entries contradicted by current observations.
We present a maintenance contract for \tool{} in which reusable experience has an explicit applicability scope, trace provenance, and a lifecycle for confirmation, revalidation, and withdrawal. The contract reserves persistence for a Reflector role and treats recalled experience as advice. The archived implementation does not establish end-to-end enforcement of every specified mechanism. The design changes retrieved context, not model weights.
The evaluation uses read-only Android workflows and archived internal comparisons. A six-task, three-repeat ready subset reports 18/18 completed runs with memory versus 3/18 without it, but a grounded single-agent control also solves the earlier hard-task slice. A separate cross-session probe is excluded from efficacy conclusions because of confounded route schedules and unvalidated state labels. We examine 13 selected real-update hazard records across nine subjects, distinguishing stale-locator observations from policy-based withholding and offline retrieval diagnostics. The latter do not measure live-device failures or actions saved. These events are UI changes, not confirmed regression bugs. The evidence supports a bounded design and case study; it does not establish general efficacy, superiority over independent GUI-agent systems, or unrestricted exploration.
\end{abstract}

\section{Introduction}

The same mobile test objective rarely appears only once. A regression check that reaches a settings page, verifies a search result, or exercises a template workflow may be rerun after a UI refresh, on a different device state, or during the next release cycle. Each run forces the tester to find the right entry point, choose an observation channel, recover from popups, and decide whether a previously valid route still reaches the intended state. LLM-based GUI agents make this repetition more visible because their interaction traces contain reusable operational facts: which shortcut reaches a target page, which screen requires OCR rather than the accessibility hierarchy, which recovery action works after a modal dialog, and which locator has become unsafe after an update. Discarding this experience forces each session to rediscover the same facts; replaying it uncritically can turn yesterday's GUI observation into today's wrong test instruction.

Recent GUI-agent research already recognizes the value of memory. Mobile and desktop agents learn from demonstrations, retrieve internal experience, maintain task memories, and evolve shortcuts or executable routines~\cite{appagent,mobileagentv2,agents,chainmemory,mobileagente,appagentx,hymem,mementogui,executablememory}; mobile GUI testing has likewise moved from random/model-based exploration toward LLM-guided and memory-augmented tools~\cite{qtypist,gptdroid,auitestagent,llmdroid,lelante,memodroid}. The testing challenge is therefore not simply adding long-term memory. It is making remembered GUI experience usable as \emph{testing evidence}: scoped to where it was observed, linked to traces that can be audited, ranked conservatively when reused, and withdrawn when app evolution contradicts it.

This paper studies whether an LLM-based mobile GUI tester can accumulate, reuse, and maintain such structured experience across repeated sessions. We use ``learn'' in the operational testing sense: retrieval-augmented experience accumulation that changes future test behavior, not model-weight updates. The goal is not to make memory always authoritative; it is to make memory useful when current evidence supports it and silent when the same evidence becomes stale.

We present a testing-evidence maintenance contract for \tool{}, motivated by a mobile GUI testing pipeline and its archived experiments. The specification combines tri-channel perception, three memory layers, explicit validity scopes, Laplace-smoothed confidence, and a four-state lifecycle (Confirmed, Tentative, Suspect, Stale). It requires that only the Reflector persist or transition entries. This is a design requirement, not a verified invariant of every archived runner; Section~\ref{sec:implementation-boundary} distinguishes the specification from the packaged implementation.

Our primary contribution is a testing-oriented validity contract for reusable experience: an entry records where it applies, why it is trusted, and when it must stop influencing actions. Memory layers, ranking, lifecycle, and writer restriction support this contract; they are not independent claims of algorithmic novelty. We contribute a design specification, an audit of its partial implementation, and a bounded empirical study. Table~\ref{tab:claim-map} maps observations to evidence and limitations. For traceability, we distinguish three aspects:
\emph{C1 (empirical characterization).} We report archived task comparisons and real-update stale-locator cases, distinguishing execution observations, scripted route comparisons, and offline policy diagnostics. The repeated-frontier slice is retained only as a protocol audit; its efficacy conclusions are withdrawn. These slices do not establish general efficacy or broad superiority over published GUI-testing systems.
\emph{C2 (testing-evidence representation).} Building on persistent-memory GUI agents, we specify scoped episodic, semantic, and procedural entries with confidence, trace links, and a four-state retrieval lifecycle (\S\ref{sec:approach}, Table~\ref{tab:lifecycle}). The ranking weights and promotion threshold are design choices; their exact implementation and use in the archived experiments are not established by the packaged memory manager.
\emph{C3 (writer-isolation requirement).} The contract requires proposed persistent changes to pass through a dedicated writer and enumerated lifecycle transitions. We state this as an implementation requirement, not a proved access-control property or an experimentally isolated contribution of the archived pipeline.

\section{Background and Motivation}

\subsection{From GUI Traces to Testing Evidence}

Modern mobile apps mix native screens, WebViews, custom-rendered widgets, map surfaces, feeds, and mini-program containers. A useful GUI trace therefore contains more than a list of taps: it records an operational claim about how a tester reached a state, which observation channel exposed the relevant element, which recovery action worked, or which route failed. In a one-off run, such claims may remain local debugging context. In repeated testing, they become candidates for reuse, and reuse changes their status: a remembered trace fact can influence the next test plan.

Once reused, a trace fact needs the discipline of testing evidence. In this paper, we treat reusable trace facts as testing evidence only when four properties are explicit. \emph{Scope} says where the fact is allowed to apply. \emph{Provenance} links the fact back to trace artifacts that can be audited. \emph{Ranking} controls how strongly the fact should influence a future plan relative to current observations. \emph{Withdrawal} prevents a previously useful fact from becoming a silent instruction after the app changes. Without these properties, memory has two symmetric failure modes: it can be too weak, forcing each run to rediscover useful facts, or too strong, replaying stale instructions with unwarranted authority.

\subsection{Motivating Example}

Consider a read-only map workflow that searches for a point of interest. During an initial run, a tester may discover that a visible search affordance is not the right entry for place search: it routes into a different search surface, and the correct place-search route is found only after backing out and selecting another entry. If this trace fact is discarded, the next session repeats the wrong route. If it is saved as a general shortcut, it can pollute neighboring workflows that happen to contain similar text. If the app later moves the entry or changes the target page markers, the same shortcut should no longer be treated as confirmed evidence.

\tool{} represents this experience as a scoped, trace-linked entry rather than as a command. The context records app and page markers, the action records the route or strategy, the outcome records whether the expected target was reached, the scope policy determines whether the entry is page-, app-, category-, or universally scoped, and temporal metadata records version and reuse history. When the same objective is tested later, the agent receives the entry as a suggestion ranked against current observations; if the current screen no longer supports it, the entry becomes Suspect or Stale and the agent must re-observe or replan.

A semantic checkpoint is an interpretable screen state associated with an objective, such as a search-results page. Seed states are checkpoints reached in the initial session. A frontier is the set of candidate next checkpoints; the controlled protocol limits additions per session. Archived label growth is only a diagnostic until screen-level validity is established.

The rest of the paper examines the design and archived evidence through five research questions; not all intended contrasts are supported by the available implementation and protocol.
RQ1 asks whether maintained experience changes task outcomes or cost; RQ2 whether it avoids repeated seed states under a repeated-session frontier protocol; RQ3 which memory layers are responsible; RQ4 whether scoped experience transfers to neighboring workflows or apps; RQ5 when the same mechanism becomes unsafe and whether maintenance mitigates it.

\section{Approach}\label{sec:approach}

This section specifies the intended maintenance contract. Its role, confidence, and lifecycle descriptions are normative design requirements, not a claim that all archived experiments executed this exact implementation. The implementation boundary is stated in Section~\ref{sec:implementation-boundary}.

In the specified design, the runtime receives a test objective, observes the screen, retrieves experience, executes bounded actions, and checks outcomes. Completed traces are required to update memory through a single write path. The archived evaluation examines selected task outcomes and maintenance cases, not enforcement or optimality of the complete design.

\begin{figure}[!t]
\centering
\scriptsize
\begin{tikzpicture}[
  >={Stealth[length=3.5pt]},
  font=\scriptsize,
  role/.style={draw,rounded corners=2pt,inner sep=1.5pt,minimum height=4mm,minimum width=10mm,align=center,font=\scriptsize\bfseries},
  mem/.style={draw,rounded corners=2pt,inner sep=1pt,minimum height=3.5mm,minimum width=13mm,align=center,font=\scriptsize,fill=blue!8},
  node distance=3mm and 3mm,
]
\node[mem] (epi) {Episodic};
\node[mem,right=of epi] (sem) {Semantic};
\node[mem,right=of sem] (pro) {Procedural};
\node[draw,dashed,fit=(epi)(sem)(pro),inner sep=2pt,rounded corners=2pt] (memstore) {};
\node[above=-0.3mm of memstore,font=\scriptsize\itshape] {Experience Memory \,\textbar\, 4-state lifecycle \,\textbar\, writer-isolated};

\node[role,below=8mm of epi,xshift=-2mm] (nav) {Navigator};
\node[role,right=of nav] (per) {Perceiver};
\node[role,right=of per] (exe) {Executor};
\node[role,right=of exe] (ver) {Verifier};
\node[role,right=of ver] (ref) {Reflector};

\foreach \a/\b in {nav/per,per/exe,exe/ver,ver/ref} \draw[->] (\a) -- (\b);

\draw[->,dashed,thick,gray!60!black] (epi.south) -- node[midway,right,font=\tiny,gray!60!black]{read} (nav.north);
\draw[->,dashed,thick,gray!60!black] (sem.south) -- (per.north);
\draw[->,dashed,thick,gray!60!black] (pro.south) -- (exe.north);
\draw[->,dashed,thick,gray!60!black] (epi.south) to[bend right=8] (ver.north);

\draw[->,line width=0.7pt,red!60!black] (ref.north) to[bend left=20] node[midway,above,font=\tiny]{write (only)} (memstore.east);
\end{tikzpicture}
\caption{Specified architecture and writer-isolation requirement. Roles read memory (dashed); the Reflector is the designated writer (bold). This diagram describes the intended contract, not verified enforcement in all archived runners.}%
\label{fig:arch}
\end{figure}
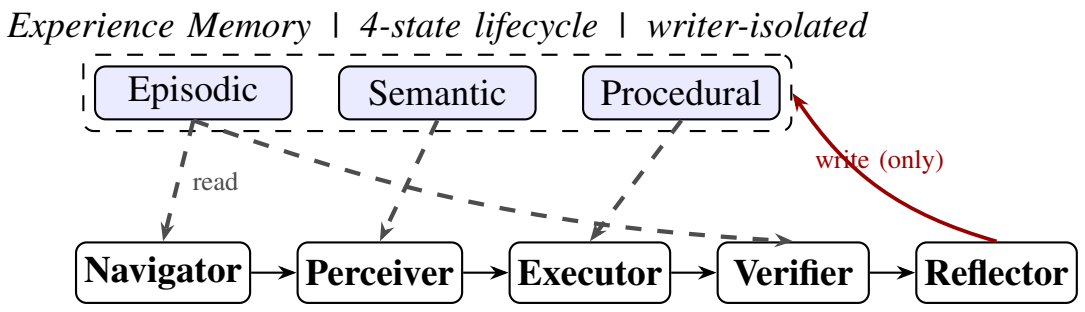

\subsection{Execution Model}

\tool{} implements mobile GUI testing as five separated roles: the Navigator plans and replans, the Perceiver observes, the Executor issues device actions, the Verifier checks postconditions and bug signals, and the Reflector distills completed traces. Agents communicate through typed payloads in a plan/execute/reflect loop. The decomposition exists to provide clean injection points for different kinds of experience (navigation, perception, recovery, verification) and to enforce that only the Reflector writes persistent memory, preventing every intermediate thought from becoming a future instruction.

\subsection{Tri-Channel Perception}

The Perceiver normalizes three observation channels into a unified element descriptor: accessibility hierarchy (cheap and structured), optical character recognition (OCR; visible text in WebView or rendered content), and visual grounding (icons or canvas-like elements without useful accessibility nodes). The default channel is chosen by a highest-success-rate lookup over the semantic memory's per-page channel-reliability entries, falling back to a target-description heuristic when no entry exists; if confidence is low, \tool{} invokes additional channels and fuses candidates.

Each perceived candidate is represented as $u=(\mathit{id}, \mathit{source}, \mathit{text}, \mathit{type}, \mathit{bbox}, \mathit{conf}, \mathit{attrs})$ where $\mathit{id}$ identifies the candidate, $\mathit{text}$ is visible or extracted text, $\mathit{type}$ is the element category, $\mathit{bbox}$ is the screen bounding box, and $\mathit{attrs}$ contains additional channel-provided attributes. $\mathit{source}\in\{\mathit{hierarchy},\mathit{ocr},\mathit{vision}\}$ and $\mathit{conf}\in[0,1]$ is channel-normalized confidence. When two channels produce overlapping bounding boxes for the same target, \tool{} merges their descriptors and keeps provenance for audit. The perception stack is a dependency of the runtime rather than the contribution of this paper; the contribution is how the resulting observations are stored, scoped, and retired as test evidence.

\subsection{Experience Representation}

We model a test session as a trace
$T=\langle(\sigma_1,a_1,o_1),\ldots,(\sigma_n,a_n,o_n)\rangle$,
where $\sigma_i$ is the observed screen state, $a_i$ is the action, and $o_i$ is the outcome.
In the design model, an experience entry is:
\begin{equation}
e=(C,A,O,V,\Theta,\Gamma)
\end{equation}
where $C$ is context, $A$ is action or strategy, $O$ is outcome, $V$ is validity scope, $\Theta=(\mathit{use},\mathit{succ},\mathit{conf})$ is reliability state, and $\Gamma=(t_{\mathit{created}},t_{\mathit{used}},\mathit{ver})$ is temporal metadata. $\Theta$ is updated by the Reflector after every reuse; $\Gamma$ is updated by the runtime. Here $\mathit{use}$ and $\mathit{succ}$ count attempts and successful reuses. Temporal fields record creation, last use, and app version; trace identifiers link the entry to its supporting observations. Splitting reliability from time avoids a circular definition: confidence is derived from $\Theta$ rather than recomputed from $\Gamma$ each retrieval.

The schema is deliberately conservative relative to general GUI-agent memory systems. A stored trajectory, shortcut, or page fact is not promoted because it looks semantically similar to the current task; it is promoted only when its scope, reuse history, and current-screen evidence jointly support reuse. This is the point at which \tool{} departs from flat trace replay or generic retrieval-augmented planning: the same remembered action can be helpful evidence inside one workflow, irrelevant in a neighboring workflow, or stale after an app update.

The validity scope $V$ is essential. Entries can be universal (e.g., dialog dismissal), category-scoped (e.g., map place-detail markers), app-scoped, or page-scoped. Scope prevents a successful route in one context from silently becoming an instruction in another. The Reflector applies $V$ via a two-step rule: (1)~if the entry references a specific package identifier or page-level locator, it receives the narrowest matching scope; (2)~otherwise, if it describes a cross-app pattern, it receives category or universal scope. Scope is persisted with app, workflow, version, and trace-provenance fields so that later retrieval can be checked against both the current query and the screen evidence that produced the entry. The runtime does not re-scope a Confirmed entry silently; it can only downgrade it to Suspect or Stale via the lifecycle.

\subsection{Three Memory Layers}

\tool{} stores entries in three layers.
Episodic memory records what happened in previous traces and supports plan recall.
Semantic memory stores versioned app knowledge such as page-transition hints, widget catalogs, and perception-channel reliability.
Procedural memory stores reusable strategies, such as recovery behaviors or non-default perception choices.
At runtime, retrieved entries are injected as suggestions into the relevant agent role, as summarized in Table~\ref{tab:memory-layers}.
They are not hard constraints: current screen observations override memory.

An episodic entry retains a completed objective, route, and outcome. A semantic entry summarizes a page fact, such as expected target markers or a previously effective observation channel. A procedural entry describes how to act, such as a recovery sequence or fallback strategy. In the WPS case below, the completed interaction is episodic evidence, the version-scoped locator-to-marker relation is semantic knowledge, and a reusable recovery strategy is procedural advice. Failure of one locator need not invalidate every higher-level strategy. Each reused entry retains its own scope and provenance.

\begin{table}[t]
\centering
\caption{Experience layers and their runtime injection targets.}%
\label{tab:memory-layers}
\scriptsize
\setlength{\tabcolsep}{3pt}
\renewcommand{\arraystretch}{1.14}
\begin{tabularx}{\columnwidth}{@{}lYY@{}}
\toprule
Layer & Stores & Injected into \\
\midrule
Episodic & Prior traces, objectives, outcomes & Navigator route examples and Verifier evidence links \\
Semantic & Versioned pages, widgets, channel reliability & Navigator page graph and Perceiver channel ranking \\
Procedural & Recovery and fallback strategies & Executor action choices and bounded recovery \\
\bottomrule
\end{tabularx}
\end{table}

The specified relevance score combines scope, semantic similarity, and recency: $\mathit{rel}(e,q)=0.4\,\mathit{scope}(V,q)+0.4\,\mathit{sim}(C,q)+0.2\,\mathit{rec}(\Gamma)$. Each component is normalized to $[0,1]$: $\mathit{scope}(V,q)$ is an ordinal compatibility score (exact page/workflow $=1.0$, app-scoped $=0.7$, category $=0.4$, universal $=0.2$, incompatible $=0$), $\mathit{sim}(C,q)$ is cosine similarity over text embeddings of the stored context and query, and $\mathit{rec}(\Gamma)=\exp(-\Delta t / \tau)$ with $\tau\!=\!30$~days. The 0.4/0.4/0.2 coefficients are fixed design priors that encode the validity contract: scope and semantic similarity receive more weight than recency. This sum is a ranking heuristic, not a logical scope guarantee: similarity and recency can still favor an incompatible entry. Applicability and current-screen checks, rather than these coefficients alone, must determine whether a suggestion can be used. They are configuration parameters, not online-learned weights.
Confidence is derived from $\Theta$ via a Laplace-smoothed success rate $\mathit{conf}(e)=(\mathit{succ}+\alpha)/(\mathit{use}+\alpha+\beta)$ with $\alpha=\beta=1$ as a neutral Beta prior; the contract permits injection of $e$ into a Confirmed slot only when $\mathit{conf}(e)\ge 0.60$, with lower-confidence entries routed to Tentative or Suspect handling. The threshold is intentionally low enough to avoid starving repeated-test reuse but paired with a distinct-session promotion rule to block one-shot confirmation. A fresh entry with $\mathit{use}{=}\mathit{succ}{=}0$ starts at $\mathit{conf}{=}0.5$ in Tentative; one successful reuse already raises confidence to $2/3$, above 0.60. Confirmation nevertheless requires success in two distinct sessions, yielding $3/4{=}0.75$. This separate guard, not the threshold alone, prevents one-shot confirmation. New procedural entries are tentative until they succeed across distinct sessions; repeated failure demotes or disables them. When retrieved entries exceed an agent's prompt budget, the design requires \tool{} to rank by relevance and confidence, preserve trace identifiers, truncate the lowest-ranked suggestions first, and fall back to current-screen planning if no verified entry remains.

\textit{Lifecycle.}
The contract requires retrieval to remain subordinate to current observations. After verification, the Reflector is required to update use counts and link entries to trace artifacts. Retrievable summaries must pass schema validation and retain provenance. These are design requirements; validation of summary structure alone does not establish provenance completeness or absence of private account information.

\subsection{Staleness Handling}

An entry becomes \emph{suspect} when version or contradiction signals fire (the app version has changed, or the Verifier emits a contradiction event), and \emph{stale} when bounded revalidation on the current screen fails. Bounded revalidation re-checks the stored markers via at most $k\!=\!2$ retries on the current screen, restricted to read-only entries; any side-effectful or login-gated entry skips revalidation and remains Suspect until the next Reflector pass. Table~\ref{tab:lifecycle} specifies the lifecycle: \textsc{Confirmed} entries are injected normally, \textsc{Tentative} with lower rank and explicit caution, \textsc{Suspect} only after current-screen validation, and \textsc{Stale} withheld from action prompts but visible to the Reflector as negative evidence. A conforming implementation must reject any state change not enumerated in the table; enforcement is not established by the archived manager.

\begin{table}[t]
\centering
\caption{Four-state lifecycle transition specification. \textsc{Conf}$=\mathit{conf}(e)$ via Laplace smoothing. \textsc{ver} compares stored vs current app version. Every transition's writer is the Reflector (required writer isolation); other roles may only \emph{propose} signals (e.g., the Verifier emits a contradiction event) but cannot persist transitions. A conforming implementation must reject unlisted transitions; this table specifies requirements, not verified enforcement.}%
\label{tab:lifecycle}
\scriptsize
\setlength{\tabcolsep}{3pt}
\renewcommand{\arraystretch}{1.12}
\begin{tabularx}{\columnwidth}{@{}llYl@{}}
\toprule
From & To & Guard condition & Writer \\
\midrule
$\bot$ & Tentative & New entry persisted from completed trace & Reflector \\
Tentative & Confirmed & $\geq\!2$ distinct-session successful reuses, \textsc{Conf}\,$\geq\!0.60$ & Reflector \\
Tentative & Stale & Repeated failed reuse ($\geq\!2$ failures, \textsc{Conf}\,$<\!0.40$) & Reflector \\
Confirmed & Suspect & \textsc{ver} bump $\vee$ Verifier contradiction event & Reflector \\
Suspect & Confirmed & Current-screen revalidation matches stored markers & Reflector \\
Suspect & Stale & Bounded replay fails on current screen & Reflector \\
Stale & --- & Withheld from injection; retained as Reflector negative evidence & --- \\
\bottomrule
\end{tabularx}
\end{table}

The contract requires a validator and writer authorization to enforce the transition table: a completed trace may create a Tentative entry; reuse updates $\Theta$ and confidence; two distinct successful reuses can promote to Confirmed; version bumps or Verifier contradictions downgrade to Suspect; current-screen validation restores Confirmed; failed bounded replay moves the entry to Stale. A conforming implementation must reject any unlisted transition or non-Reflector write. The archived implementation does not establish these checks.

\subsection{Running Example: A Version-Scoped Locator}
The WPS version pair changes from 14.43 to 14.45. The real-update replay shows an old template-category coordinate pointing to a ledger-details chip, although the broader template entry remains available. A remembered coordinate and the intended target must therefore be checked separately.

To illustrate the lifecycle, consider an entry distilled from the v1 route. The Reflector stores it as Tentative, with its locator, expected marker, version scope, and trace. Successful reuse in two later sessions permits confirmation. A version change triggers Suspect handling; failed bounded revalidation leads to Stale handling and withdrawal from action prompts, while retaining the trace for diagnosis. The coordinate mismatch is recorded RQ5 evidence. The creation and promotion sequence here illustrates the specified lifecycle; it is not an additional measured end-to-end trajectory.

\subsection{Implementation and Provenance Hooks}\label{sec:implementation-boundary}

The packaged memory manager implements JSON persistence, recall, recording, and consolidation. Its scoring function adds 100 for case-ID matches, 50 for title matches, 8 for matching tags, and 20 for matching mock keys or entry hints, with an additional exact-memory-ID bonus in recall. This is not the normalized scope/similarity/recency formula specified above. The inspected manager and schemas do not establish the four named lifecycle states, the specified confidence/promotion rule, or authenticated Reflector-only writes. A record command is exposed without a writer-role argument. An external orchestration layer could impose additional constraints, but the packaged evidence does not verify that boundary. We therefore do not attribute measured outcomes to enforcement of the full contract.

The packaged pipeline uses profiles and a Model Context Protocol (MCP) device-tool interface. Archived JSON records contain tool actions, outcome summaries, and evidence paths; some paths refer to files not included in the review subset. These records aid inspection but do not by themselves prove target correctness, complete screenshot coverage, or implementation of the specified lifecycle. Stored experience and tool records must be interpreted according to their actual fields rather than the idealized contract.

The MCP server is the device-tool boundary, not a memory learner. In the intended role design, the Navigator proposes a route; the Perceiver obtains observations; the Executor issues actions; the Verifier checks the resulting state; and the Reflector receives the completed trace for a validated memory update. Tool observations and agent proposals remain distinguishable so that an intended action is not mistaken for an observed success.

The design is not tied to a particular LLM provider. Role prompts request typed proposals, while backend-specific configuration is externalized. Validating JSON shape is not equivalent to enforcing writer identity or lifecycle transitions; those properties need dedicated implementation and tests. Evaluation-specific backbone, baseline, and artifact boundaries are reported in \S\ref{sec:eval}.

\section{Evaluation Design}\label{sec:eval}

Table~\ref{tab:claim-map} maps each contribution to the evaluation slice that can support it; result counts are reported only in the corresponding RQ subsections.

\begin{table}[t]
\centering
\caption{Claim-to-evidence design map. Each row states the supporting slice and the boundary of the claim; concrete counts are reported in the corresponding results subsection.}%
\label{tab:claim-map}
\scriptsize
\setlength{\tabcolsep}{3pt}
\renewcommand{\arraystretch}{1.12}
\begin{tabularx}{\columnwidth}{@{}L{0.22\columnwidth}YY@{}}
\toprule
Claim & Supporting slice & Evaluation boundary \\
\midrule
Archived task outcomes & RQ1 executable context and internal comparisons & Descriptive runtime outcomes; architecture attribution is limited \\
Protocol audit & RQ2 archived design and label checks & No efficacy conclusion; excluded from coverage and architecture claims \\
Layer sensitivity & RQ3 hidden-from-retrieval ablations & Outcomes depend on the selected harness and task pool \\
Route reuse & RQ4 recall and scripted route comparisons & Added cold-route probes confound acceleration attribution \\
Stale-locator hazards & RQ5 selected real-update cases and policy diagnostics & Separate observed changes from inferred prevention and retrieval hits \\
\bottomrule
\end{tabularx}
\end{table}

The evaluation examines experience reuse and maintenance within one runtime family. Third-party tools provide executable context and reproduction status. In the intended control design, Full vs NoMem removes Experience Memory, SA is a stateless monolithic controller, SA+Mem adds recalled entries, and Maint.-off disables stale/version-scope maintenance. Actual implementation and oracle limitations must be assessed per slice: RQ2 does not establish these intended contrasts and is excluded from efficacy claims. RQ5 examines stale-change triage under the recorded replay controls.
The archive inventories 158 runs, 493 in-window logcat files, 120 RQ2 state records, 37 persisted scope records, 280 sensitivity checks, and 20 adapted LLMDroid context runs with 4,024 events. The 158-run inventory includes RQ2 and is not a denominator of validated efficacy trials. We use slice-specific denominators for each retained comparison.
RQ5 studies stale-experience handling. RQ2 documents why the archived cross-session probe cannot answer its intended question. RQ1 calibrates solvability; RQ3 and RQ4 are supporting analyses. Identifiers are retained for traceability to the archived packets:
\begin{itemize}[leftmargin=*]
\item \textbf{RQ1:} How does memory affect task completion, action count, and execution time under same-backbone controls?
\item \textbf{RQ2:} Does prior-session evidence persist as new safe checkpoints under a controlled repeated-frontier protocol?
\item \textbf{RQ3:} Which memory layers matter on the harder task pool?
\item \textbf{RQ4:} Does scoped experience help the evaluated neighboring workflows and limited cross-app probes?
\item \textbf{RQ5:} When does accumulated experience become stale, and does maintenance mitigate it?
\end{itemize}

\textbf{Naming.}
We use ten short labels: \textsc{Full} = \tool{} all layers + stale/version filtering; \textsc{NoMem} = Experience Memory disabled; \textsc{Maint.-off} = \tool{} with stale/version filtering disabled (other layers active); \textsc{SA} = same-backbone monolithic SingleAgent, memory disabled; \textsc{SA+Mem} = SA with the same Experience Memory as Full; \textsc{NoE}/\textsc{NoS}/\textsc{NoP} = Full with the Episodic/Semantic/Procedural layer hidden from retrieval (RQ3); \textsc{FlatReplay-lite} = Full with flat replay instead of scoped semantic validation; \textsc{StrippedFlatRAG} = internal flat-RAG ablation of \tool{}'s retrieval layer (RQ5 control). External baselines (Monkey, LLMDroid) retain tool names.

\textbf{Subjects and tasks.}
The counted Android subjects cover finance/super-app, search, map/navigation, productivity, health, travel, social/content, video/content, and utility apps: Alipay, Baidu Search, Amap, WPS Office, DXY Doctor, an anonymized commercial health-companion app, Douban, Ctrip, Bilibili, and the Fossify suite. Task-level experiments use non-destructive objectives such as reaching weather, search-result, place-detail, document-list, health-service, or content-tab surfaces; no payment, order, account mutation, file upload, or appointment action is counted. Counted task pools, version pairs, and replay events are fixed by whitelist before analysis; setup failures and diagnostic attempts remain in the artifact but do not alter denominators. The submission subject/evidence inventory and representative task list are recorded in \texttt{outputs/subject\_app\_inventory\_20260513.*} and \texttt{ARTIFACT\_MANIFEST.md}. Android activity counts are treated as metadata because hybrid and feed-style apps expose many semantic states inside one Android activity.

\textbf{Baselines and control conditions.}
The evaluation uses two kinds of comparison, with different evidentiary roles. External rows provide executable context for random exploration and public LLM-guided GUI testing; internal rows are the counted mechanism controls.
Monkey and LLMDroid are executed end-to-end as app-level executable-context rows; an additional Monkey smoke checks the Baidu Search version pair used in RQ5.
The LLMDroid condition follows the public LLMDroid recipe~\cite{llmdroid} with harness compatibility patches applied to the upstream artifact, and reaches the same hosted multimodal backbone as \tool{} through an OpenAI-compatible endpoint (\texttt{configs/backbone.yaml}); it is an adapted executable datapoint, not a claim of published-system superiority.
SA and SA+Mem are the archived labels for intended single-controller comparisons without and with memory. Their aggregate outcomes alone do not verify controller architecture. The RQ1 SA+Mem wrapper registers an all-layer profile and calls a shared hard-pool runner before adding architecture metadata; its wrapper is not independent evidence that role separation was disabled in the underlying runtime. We retain its reported outcomes descriptively, not as proof of architecture attribution.
For additional related systems, a P1 official-baseline mini-slice reports reproduction status and boundary evidence rather than broad ranking. MemoDroid's public memory pipeline runs on its demo and on a Route-A adapter over four RQ5 v1 traces (11 segments, 12 retrieval queries, hit@1 $=9/12$); LLMDroid's public DroidBot module supplies the executable row above; GPTDroid is blocked by legacy Appium/Python/OpenAI setup; and DroidBot is blocked at accessibility-service activation. Blocked attempts are not scored as negative baseline results.
We then use same-backbone protocol controls for mechanisms that third-party baselines do not isolate directly: dynamic memory without stale/version-scope maintenance, and flat replay without scoped semantic validation.
Together, the executable baseline rows and internal-control rows cover random exploration, an adapted LLM-guided testing workflow (LLMDroid), same-backbone stateless control (SA), dynamic memory without stale/version-scope maintenance (Maint.-off), and flat replay.

\textbf{Stripped flat-RAG diagnostic.}
\textsc{StrippedFlatRAG} retrieves segments reconstructed from the audited RQ5 records. It is offline, not a fresh device execution or faithful MemoDroid reproduction. Its runner sets \texttt{stale\_fail} equal to a stale-segment match and assigns waste as one action for Camera and two for other core events. These derived fields are not observed failures or measured savings. Queries include the audited stale locator and observed v2 behavior, so this is a retrospective diagnostic rather than a prospective retrieval benchmark. Table~\ref{tab:rq5-replay} separates this evidence from update observations.

\textbf{Metrics.}
We report target readiness or pass/block outcomes for task-level workflows; semantic UI states for repeated app-level exploration; action and wall-clock proxies for cost; target-package crash/ANR markers for conservative bug evidence; and experience recall, stale detection, and wasted stale attempts for memory behavior.
The archived RQ2 signature is a triple $(\mathit{pkg}, \mathit{label}, \mathit{fg})$, comprising the target package, a protocol-defined checkpoint label saved in the run result, and the foreground Android component. Two signatures differ if any field differs. A label alone is not evidence that its semantic objective was reached. Scheduled checkpoints specify the next-session targets, capped as described in \S\ref{sec:rq2}. Until labels are independently adjudicated against screenshots and traces, this metric measures protocol-label accumulation, not verified semantic-state growth or open-ended exploration.
Functional bug candidates enter the bug-yield count only after artifact audit.
Crash or ANR evidence is counted when the logcat marker names the target package within the audited execution window; harness, accessibility, or UiAutomator process failures remain diagnostic.
A controlled oracle sanity supplement accepts 3/3 true states and rejects 3/3 seeded false expectations, supporting assertion discipline without changing bug yield.

\textbf{Protocol and statistics.}
Task-level comparisons use paired configurations whenever possible. For binary paired outcomes we use an exact paired sign test or the equivalent McNemar/binomial test when only discordant pairs matter; for paired continuous metrics we use Wilcoxon signed-rank tests and report the paired effect statistic. Repeated-session probes preserve Experience Memory for Full and reset or disable it for NoMem. Unless otherwise stated, the counted task-level repeats use the same Android 16 USB device and the same hosted LLM backbone across \tool{} and same-backbone baselines.

\textbf{Artifact boundary.}
The archived evaluation package contains \texttt{ARTIFACT\_MANIFEST.md} and an index at \texttt{outputs/ARTIFACT\_MANIFEST.md}, mapping tables to traces, summaries, and scripts. No public artifact endpoint is currently available; references to archive files identify local evidence, not publicly accessible replication resources. Diagnostic attempts remain separate from counted runs. The RQ5 core contains 13 events over nine subjects; repeated Notes events are not independent subjects. Numerical experiment tables are retained from the archived evaluation, not newly rerun experiments.

\section{Results}

The results follow the claim-to-evidence map in Table~\ref{tab:claim-map}. We first summarize runnable context and same-backbone mechanism slices, then discuss the repeated-frontier protocol audit and cross-version stale-locator evidence. RQ-specific tables use fixed denominators, while the artifact supplies the broader run/log/state/scope footprint.
The core empirical question is not whether one controller can finish one task once, but whether maintained experience changes repeated testing behavior while remaining safe to reuse.
RQ1 first calibrates this point: pass rate saturates on low-risk tasks, and the harder pool separates route families without making task solvability the headline.
RQ5 examines selected stale-locator cases and distinguishes observations from policy and retrieval diagnostics. RQ2 is retained as a protocol audit with no efficacy conclusion: the archived design and state labels do not establish successful frontier persistence. RQ3 reports layer-sensitive outcomes, and RQ4 describes scripted route comparisons rather than general transfer efficacy.

\subsection{RQ1: Separating One-Shot Solvability from Maintained Experience}

RQ1 provides calibration slices for one-shot outcomes, current-screen grounding, and recalled experience. External-tool rows give runnable context. Full/NoMem and SA/SA+Mem are archived internal comparison labels; their status counts describe these runs, but the implementation boundary prevents treating them as a verified isolation of the full specified lifecycle or role architecture.

\begin{table}[t]
\centering
\caption{Executable app-level context subset. Cov.\ is the fraction of manifest-declared activities visited within each run; Markers is target-package crash/ANR markers. The adapted LLMDroid row uses MobileFlow cloud Android devices (Android 9--15; replacements for ADB stability). Cross-version replay results are in Table~\ref{tab:rq5-replay}; the withdrawn RQ2 comparison is discussed in Section~\ref{sec:rq2}.}%
\label{tab:rq1-baselines}
\scriptsize
\setlength{\tabcolsep}{2pt}
\begin{tabular*}{\columnwidth}{@{\extracolsep{\fill}}lrrrr@{}}
\toprule
Tool & Runs & Act./app & Cov. & Markers \\
\midrule
Monkey & 18 & 4.39 & 1.24\% & 0 \\
LLMDroid & 20 & 8.25 & 0.95\% & 12 \\
\midrule
Monkey (Baidu v2) & 1 & 3.00 & 2.33\% & 0 \\
\bottomrule
\end{tabular*}
\end{table}

Table~\ref{tab:rq1-baselines} records executable app-level context. The completed Monkey rows show random-event activity reach and crash/ANR markers. The adapted LLMDroid row adds 20 valid runs (16 full-window) with 4,024 total events, 8.25 reached activities per run, 0.9535\% mean activity coverage (0.9066\% over full-window runs only), and 12 target-package crash/ANR markers. These rows establish runnable context; the mechanism claims are tested in the same-backbone slices below.

The low-risk task pool (7 task-level objectives $\times$ 3 paired repeats = 21 paired runs) saturates: both Full and NoMem pass all 21, but Full uses fewer actions and less time (Wilcoxon $p=0.0313$ and $p=0.0157$). The harder task pool removes the ceiling. Table~\ref{tab:hard24} reports the frozen Hard-24 rerun: the paper-safe ready subset covers six pre-ready tasks $\times$ three repeats, where Full passes 18/18 and NoMem passes 3/18 (15 Full-only pairs, two-sided exact paired binomial $p=0.000061$). The retained full matrix is Full 18/24 and NoMem 3/24; the remaining six cells per condition correspond to two task definitions that the frozen manifest had already pre-flagged as requiring targeted repair.

\begin{table}[t]
\centering
\caption{Archived Hard-24 Full-vs-NoMem summary. The ready subset excludes pre-flagged repair tasks. The paired $p$ value is run-level and assumes independent pairs; repeats share task definitions.}%
\label{tab:hard24}
\scriptsize
\setlength{\tabcolsep}{3pt}
\begin{tabular*}{\columnwidth}{@{\extracolsep{\fill}}lrrr@{}}
\toprule
View & Full & NoMem & Paired $p$ \\
\midrule
Ready subset & 18/18 & 3/18 & 0.000061 \\
Full retained matrix & 18/24 & 3/24 & 0.000061 \\
\bottomrule
\end{tabular*}
\end{table}

SA, SA+Mem, and Maint.-off also solve this pool under different assumptions: live current-screen grounding (SA), explicit recalled experience in a monolithic controller (SA+Mem), and valid dynamic-memory hints without stale/version-scope filtering (Maint.-off).
Table~\ref{tab:rq1} keeps the earlier 12-cell mechanism boundary so that SA, SA+Mem, Maint.-off, and layer ablations remain comparable under one task slice. In that slice, SA uses fewer trace steps, but longer time and more current-screen perception.
An offline cost-field audit aggregates run-level statistics over 158 counted runs (RQ1+RQ2+RQ3+RQ4+RQ5). Because provider prompt/completion/image-token fields were not logged uniformly, the paper reports replayable tool-call and trace-text proxies.

\begin{table}[t]
\centering
\caption{Same-backbone mechanism boundary on the original 12-cell harder task slice (per-run averages: trace Steps, wall-clock Time, Screenshots, direct Grounding calls, Element checks).}%
\label{tab:rq1}
\scriptsize
\setlength{\tabcolsep}{2pt}
\begin{tabular}{lrrrrrr}
\toprule
Condition & Pass & Steps & Time & Shot & Ground & Elem \\
\midrule
\tool{} Full & 12/12 & 29.00 & 41.54s & 3.50 & 0.00 & 16.50 \\
Maint.-off & 12/12 & 29.00 & 40.82s & 3.50 & 0.00 & 17.25 \\
\tool{}-NoMem & 0/12 & 23.25 & 33.71s & 3.50 & 0.00 & 11.75 \\
SA & 12/12 & 18.08 & 46.93s & 6.00 & 2.50 & 0.00 \\
SA+Mem & 12/12 & 29.00 & 41.60s & 3.50 & 0.00 & 17.25 \\
\bottomrule
\end{tabular}
\end{table}

Table~\ref{tab:rq1} separates solvability from cost.
NoMem's lower Steps and Time reflect early termination because it blocks on 12/12 in this mechanism slice.
SA's lower Steps reflect a perception trade-off, visible as Shot=6.00 vs 3.50 and Ground=2.50 vs 0.00; adding memory to the monolithic controller (SA+Mem) returns the same no-grounding route family as Full.
Maint.-off matches Full on non-stale tasks because both can use valid dynamic-memory hints; RQ5 then isolates the maintenance benefit when stale entries arise.

On the Alipay China Weather supplement, both conditions pass 3/3; per-run cost details are in the artifact.

\textbf{Framework continuity evidence.}
The strict in-window crash/ANR audit scans 493 logcat files and retains one non-deterministic target-package ANR candidate on a single Alipay search activity, with 0 confirmed crashes after follow-up rechecks. This log-audit slice checks harness continuity and assertion discipline; the app-evolution safety result appears in RQ5 (\S\ref{sec:rq5}), where maintained memory surfaces 13 cross-version stale-change candidates that Maint.-off silently replays. Diagnostic crash-reproduction attempts and not-counted candidates are indexed in \texttt{ARTIFACT\_MANIFEST.md}.

\textbf{Finding 1.}
RQ1 separates the factors that are easy to conflate in GUI-agent evaluation. The frozen Hard-24 ready subset expands the Full-vs-NoMem denominator from 12 to 18 cells per condition and keeps the same qualitative boundary: Experience Memory changes the ready subset from 3/18 to 18/18. The 12-cell mechanism slice then shows that adding the same memory to SA recovers the Full route family. Live-grounded SA shows that strong current-screen grounding can solve one-shot tasks; Maint.-off shows that maintenance is not needed before entries become stale. These outcomes do not establish cross-session semantic progress. RQ5 separately examines stale-memory withdrawal; the RQ2 protocol does not support the intended persistence claim.

\subsection{RQ2: Protocol Audit Without an Efficacy Conclusion}\label{sec:rq2}

The archived five-session probe covers Baidu Search, Amap, WPS Office, and Douban. We withdraw its semantic-coverage and frontier-persistence conclusions for three reasons.

First, the selection metadata specifies a safe-frontier schedule for Full while NoMem repeats the generic seed route. A shared cap on checkpoints does not make these policies equivalent: a difference in accumulated labels can arise from the assigned schedule rather than successful experience use. The existing comparison cannot isolate this effect.

Second, the state signature includes a protocol-defined checkpoint label. The guarded CSV contains 120 records, whereas the visual subset packages 26 representative screenshots, not 120 independently inspected observations. The visual manifest states that the guarded runs did not persist XML hierarchy dumps. Amap screenshots labeled \texttt{poi\_results}, \texttt{poi\_detail}, and \texttt{map\_layer\_panel} do not clearly show the claimed result list, detail panel, or opened layer panel. Target-package checks and distinct assigned labels therefore do not establish achieved semantic states.

Third, the archived SA+Mem wrapper calls the Full session function and then rewrites the condition and architecture metadata. That wrapper alone does not demonstrate an independently implemented single-agent controller with role separation disabled. We do not use this RQ2 control to attribute outcomes to memory or architecture.

The archived count table and supplemental comparisons are excluded from the efficacy results. Their original records remain preserved for audit; no replacement semantic-state totals are inferred. The earlier foreign-package record \texttt{RQ2-IAA-004} is absent from the guarded CSV, but resolving that lineage issue does not repair these design and measurement limitations.

\textbf{RQ2 outcome.} The intended research question remains unanswered. A new test must give every condition the same candidate information and decision policy except for the memory intervention, verify controller configuration at execution time, and judge reached states independently of planned checkpoint labels. The original traces and runtime verdicts cannot substitute for that evaluation.

\subsection{RQ3: Memory-Layer Contributions}

Table~\ref{tab:rq3} ablates the three memory layers on the harder task pool while keeping the agent decomposition and perception stack fixed.
In every ablated variant the Reflector still runs and writes summaries, but the corresponding layer is hidden from retrieval, so the differences below isolate retrieval-time use rather than reflection cost.
A task-selection audit freezes the Baidu/Amap/WPS/Douban objectives before comparison, records the non-destructive target-surface criteria, and indexes supplementary or out-of-scope candidates in the artifact.
Full passes all 12 runs.
Removing semantic memory, procedural memory, or all explicit memory drops pass rate to 0/12; removing episodic memory drops pass rate to 6/12.

\begin{table}[t]
\centering
\caption{RQ3 memory-layer ablation on the harder task pool. Wins/Ties counts paired runs where Full passes and the variant fails (Wins) versus where both pass (Ties); the sign-test $p$ value compares each variant against Full on the 12 paired runs.}%
\label{tab:rq3}
\scriptsize
\setlength{\tabcolsep}{2.5pt}
\begin{tabular}{lccrrrr}
\toprule
Variant & Layers & Pass & Acts & Time & Wins/Ties & $p$ \\
\midrule
Full & E+S+P & 12/12 & 28.25 & 39.65s & --- & --- \\
NoE & S+P & 6/12 & 25.50 & 36.43s & 6/6 & 0.03125 \\
NoS & E+P & 0/12 & 23.00 & 33.55s & 12/0 & 0.000488 \\
NoP & E+S & 0/12 & 21.00 & 30.86s & 12/0 & 0.000488 \\
NoMem & none & 0/12 & 21.00 & 31.28s & 12/0 & 0.000488 \\
\bottomrule
\end{tabular}
\end{table}

The table retains archived run-level paired tests, with the three 12/12-vs-0/12 contrasts reaching the paired-sign floor at $n\!=\!12$ ($p\!=\!0.000488$). Repeats share four task definitions, so these values do not establish significance across independent tasks or apps; the four contrasts are also exploratory and unadjusted for multiplicity. As in Table~\ref{tab:rq1}, lower action and time counts for ablated variants reflect early blocking, not lower cost for successful testing.

NoE's 6/12 passes are informative: they occur on simpler tab/list workflows where semantic and procedural hints are enough to select and validate the target, while Baidu and Amap tasks in the harder pool need episodic routes more often because the agent must distinguish the remembered search or detail path from visually similar alternatives. NoS and NoP fail all runs for different reasons: NoS, replicated by a FlatReplay-lite control, lacks scoped semantic validation, while NoP lacks executable fallback behavior. The NoS/NoP/NoMem rows all saturate at 0/12 because every task in the hard pool requires checkpoint-marker validation \emph{and} fallback recovery on at least one screen; a pool with weaker oracle requirements would attenuate this saturation.

\textbf{Finding 3.}
The archived outcomes are sensitive to the layer profiles in this harness and selected task pool. They are consistent with the intended semantic-validation and procedural-fallback roles, but do not establish that three separate layers are necessary or that simpler representations cannot achieve the same outcomes. Repeated runs and task-specific oracle requirements limit the generality of the comparison.

\subsection{RQ4: Scoped Transfer and Adjacent-Entry Selection}\label{sec:rq4}

Transfer is useful only under bounded conditions.
In a seven-task leave-one-out check over the representative non-destructive task objectives (each held out once while the remaining six serve as source tasks) with three repeats per held-out task (21 runs), source-task experience is recalled without target leakage. This slice is designed as a recall-fidelity check; acceleration is evaluated in the controlled extension below.
A controlled transfer-acceleration extension adds four generic exploratory probes to the cold condition (dismiss-permission-dialog, open-search-bar, switch-tab, scroll-to-bottom) and allows transfer-guided runs to skip them when a matching procedural entry is available.
Both conditions pass 15/15 runs, but transfer-guided execution reduces actions from 28.8 to 24.8 per run; the paired Wilcoxon statistic is reported in the analysis artifact.

Entry-selection supplements test adjacent shortcuts or tabs.
Across formal Ctrip, Douban, Baidu Map, and WPS checks, cold exploration reaches 21/24 targets, triggers 24/24 wrong-route or wrong-tab attempts, and averages 6.0 actions.
Memory-guided execution reaches 24/24 targets, triggers 0/24 wrong attempts, and averages 3.0 actions.
The archived Wilcoxon value ($p=4.8\times10^{-7}$) and Cliff's $\delta \approx -1.0$ describe these route-level counts. They do not establish causal acceleration because the cold policy deliberately includes a neighboring-route attempt, and repeated executions are not independent task samples.

\begin{table}[t]
\centering
\caption{RQ4 archived recall and scripted route diagnostics. The cold extension deliberately adds four probes; its action difference is not an unbiased transfer-speed estimate.}%
\label{tab:rq4}
\scriptsize
\setlength{\tabcolsep}{2.2pt}
\begin{tabular}{lrrrr}
\toprule
Check & Runs & Pass & Recall/run & Actions \\
\midrule
Leave-one-out transfer & 21 & 21/21 & 5.0 & 10.9 \\
Cold controlled extension & 15 & 15/15 & 0 & 28.8 \\
Transfer-guided extension & 15 & 15/15 & 5.0 & 24.8 \\
\bottomrule
\end{tabular}
\end{table}

Table~\ref{tab:rq4} separates recall from route accounting. The controlled extension adds four probes only to the cold route, matching the four-action mean difference. This construction illustrates skipping known probes but cannot establish autonomous acceleration. The entry-selection supplement similarly assigns an initial neighboring shortcut to the cold policy, so its wrong-attempt difference is partly built into the intervention. Neither comparison establishes broad cross-app transfer.

\begin{table}[t]
\centering
\caption{RQ4 entry-selection supplement. Wrong attempts are plausible neighboring shortcuts or tabs tried before the target.}%
\label{tab:rq4-entry}
\scriptsize
\setlength{\tabcolsep}{2pt}
\begin{tabular}{lrrrr}
\toprule
Condition & Runs & Ready & Wrong att. & Actions \\
\midrule
Cold exploratory & 24 & 21/24 & 24/24 & 6.0 \\
Memory-guided & 24 & 24/24 & 0/24 & 3.0 \\
\bottomrule
\end{tabular}
\end{table}

Table~\ref{tab:rq4-entry} aggregates formal entry-selection checks across Ctrip, Douban, Baidu Map, and WPS. The cold exploratory route first validates one plausible neighboring shortcut before correcting to the target; memory-guided execution uses a scoped entry that names the target shortcut or tab and the semantic marker expected after the tap, eliminating 24/24 wrong attempts and improving readiness from 21/24 to 24/24. This focused slice tests scoped transfer at the point where wrong-route attempts occur: a local, auditable decision boundary.

\textit{Mini case study (Ctrip, A8).} To reach the train-ticket booking surface from the Ctrip home screen, cold runs first tap the adjacent ``Hotel'' chip before correcting to ``Train''; the page-scoped memory entry names the ``Train'' chip and the expected departure/arrival marker. Memory-guided runs verify that marker before commit and skip the wrong-route tap; the same local adjacent-entry pattern appears on Douban.

\textbf{Finding 4.}
The archive reports retrieval on the selected leave-one-out setup and records the costs of prescribed routes. The cold-route additions and prescribed neighboring-entry attempt prevent an unbiased estimate of acceleration or wrong-route avoidance. Generalization requires matched decision policies, equal information except for recalled experience, and held-out workflows with independently checked target outcomes.

\subsection{RQ5: Real-Update Cases and Stale-Entry Diagnostics}\label{sec:rq5}

RQ5 examines whether historical locators can become invalid after real APK updates and how scope checks would handle such entries. We distinguish observed stale-locator cases, archived policy summaries, and offline retrieval diagnostics. They are not interchangeable forms of evidence.

The core triage packet contains 13 records across nine packages: seven open-source utilities and two commercial apps. Nine records are Fossify settings-label changes across six packages, including four Notes records; the other four concern File Manager's editor toolbar, Camera's control swap, Baidu's onboarding gate, and a WPS template-category coordinate. The packet labels these as intended or benign revisions, not confirmed unintended regressions. Repeated Notes records and a shared settings-change family do not constitute independent samples of update diversity.

The archived no-detection summary reports 13 stale attempts and a 25-action accounting total. We do not interpret that total as 25 independently measured saved actions. In particular, the WPS source explicitly distinguishes a minimum of one observed low-level waste action from a two-action paper-accounting value. The commercial and Camera source summaries describe version-scope withholding as a mitigation that \emph{should} occur; later consolidation records mark all four extension events as withheld. This difference in provenance does not establish a separately executed Full condition for every event. We therefore retain the cases but do not claim a verified 13/13 prevention rate.

\begin{table}[t]
\centering
\caption{RQ5 evidence types. Counts describe archived records, not interchangeable independent device trials.}\label{tab:rq5-replay}
\scriptsize
\setlength{\tabcolsep}{3pt}
\begin{tabularx}{\columnwidth}{@{}L{0.25\columnwidth}YY@{}}
\toprule
Evidence & Archived scope & Supported interpretation \\
\midrule
Real-update triage & 13 records, 9 packages & Selected stale-locator cases; no confirmed regression bugs \\
No-detection summary & 13 attempts; 25 action units & Archived accounting, not a verified savings estimate \\
Full guard summaries & Policy decisions and consolidated withholding fields & Per-event live execution provenance is not established uniformly \\
Flat retrieval & 13/13 core stale-segment hits & Retrospective retrieval, not measured device failure or prevention \\
Breadth extension & 4 core non-settings cases plus Voice Recorder & Overlaps the core; not five additional independent cases \\
\bottomrule
\end{tabularx}
\end{table}

The offline FlatRAG runner retrieves from reconstructed event records and converts matches into failure flags without executing the retrieved instructions on a device. Its query already includes stale-locator and v2-observation information from the audit. The reported 13/13 core hits and derived failure totals therefore show what is retrieved under that construction, not prospective failure rates or a matched comparison with an executing agent. The lexical Voice Recorder extension has the same interpretive boundary. Results attributed to token-overlap and sentence-transformer backends remain offline diagnostics, not independent replications of device outcomes.

\textit{Case evidence.} In Camera 1.2.0 to 1.3.0, hierarchy records place the old switch-camera coordinate inside the new media-preview control, and the source reports a resolver opening after the tap. In WPS 14.43.0 to 14.45.0, the old category coordinate selects ledger templates rather than the prior summary-report category; this does not imply that template navigation as a whole broke. Baidu 15.57.1.10 to 15.60.0.10 introduces an onboarding gate at the old search entry. File Manager's editor exposes the replacement Save as route through an overflow menu. These examples motivate current-screen revalidation but do not establish the false-withdrawal rate or the cost of reacquiring still-valid experience.

A separate controlled stale-entry diagnostic assigns outdated routes to an unguarded condition and corrected validation/recovery routes to a guarded condition. Its archived summaries report 0/15 versus 15/15 readiness over five tasks. Because this diagnostic supplies the route interventions, it demonstrates behavior on chosen stale inputs, not autonomous stale-change discovery or general recovery accuracy. Version-scope candidate checks likewise flag changes in version, not independently verified defects.

\textbf{Finding 5.}
The selected cases show concrete ways in which earlier GUI locators can cease to denote the intended target after updates. Scope checks provide a conservative reason to withhold or revalidate them. The archive does not uniformly verify end-to-end prevention across all events, and the offline retrieval diagnostic cannot fill that gap. A prospective study with execution traces for both conditions, unchanged-locator controls, and revalidation-cost measurements is required to establish net benefit.

\section{Threats to Validity}\label{sec:threats}

\textbf{Construct and conclusion validity.}
The measures target repeated GUI testing rather than generic exploration. Readiness, step counts, and stale-attempt metrics describe different aspects of the selected workflows; the RQ2 semantic-state measure is withdrawn because label assignment is not an independent oracle. Cost uses wall-clock, tool-call, screenshot, grounding, and element-check proxies in the archived 158-run audit, which includes the withdrawn RQ2 slice and must not be read as 158 validated efficacy trials. The artifact contains 280 local parameter-sensitivity checks, 120 RQ2 metadata records, and 37 persisted scope records. These inventories do not establish optimal parameters, semantic correctness, or cross-backbone robustness.

The annotation packets require a narrower interpretation than the original manuscript. A June 5 report identifies the B pass as a Codex machine-side cross-check, not a second human annotation. A later completion report and canonical packets revise labels without establishing an independent human protocol. We therefore do not report two-human inter-rater reliability or treat B labels as human ground truth. Recalculation on the canonical packets gives 59/60 exact agreement for each semantic-state field, 37/37 for scope, 13/13 for hazard validity and waste checks, and 11/13 for revision subtype. Hazard validity and waste fields have zero label variance; Cohen's $\kappa$ is undefined, not 1.000. Scope contains two labels (24 and 13 records), so its descriptive $\kappa=1.000$ is defined. The semantic fields and subtype yield descriptive A/B $\kappa$ values of approximately 0.982, 0.964, and 0.581, respectively, without establishing annotator independence.

One disagreement, \texttt{RQ2-IAA-004}, concerns an Alipay foreground component labeled as an Amap result in earlier May 18/19 runs. Its state ID is absent from the later 120-row guarded CSV, whose Full/NoMem totals can be reproduced without package/component mismatches. This excludes that particular record from those totals, but does not establish their semantic correctness: Amap screenshot spot-checks do not clearly substantiate several assigned checkpoint labels. The protocol-defined labels require independent adjudication before RQ2 can support a semantic-coverage claim.
Sign-test $p$ values at $n\!=\!12$ saturate at the paired-sign floor, so we interpret statistical results together with paired traces and artifact-level evidence.

\textbf{Internal validity.}
The design has the following attribution limits.
(i)~Paired runs and foreground guards reduce some execution variation; the manifest makes selection inspectable. They do not eliminate LLM nondeterminism or make repeated runs on the same task independent. Run-level tests do not establish population-level generalization across apps.
(ii)~RQ2 combines condition-dependent route schedules with protocol-defined state labels. Its SA+Mem wrapper does not independently establish the claimed controller difference. The less-scripted and random-frontier supplements do not repair these concerns; RQ2 is excluded from efficacy conclusions.
(iii)~The archived hard-pool audit records the original counted objectives, the frozen Hard-24 manifest, non-destructive target-surface criteria, supplementary candidates, and exclusion reasons. The Hard-24 rerun retains the full 24-cell matrix per condition, but the primary paper update uses the 18-cell ready subset because two task definitions were pre-flagged for repair before execution. This audit makes task selection inspectable and separates counted ready cells from pre-flagged repair tasks, login walls, unsafe/order flows, unstable hierarchies, OCR/coordinate-heavy probes, transport instability, or non-discriminative candidates.

\textbf{External validity and ethics.}
The evidence targets repeated, read-only Android GUI testing through complementary slices rather than a single benchmark table: task-level paired runs, 4 apps $\times$ 5 frontier sessions, transfer and ablation probes, a 13-event FlatRAG-bracketed RQ5 core, and a 5-row non-settings breadth view across 10 behavior-level subjects and 6 UI-change families. Nine RQ5 core rows share a Fossify settings-label family, while the breadth rows add Camera, File Manager, Voice Recorder, and two commercial probes. We also ran Monkey and an adapted LLMDroid workflow end-to-end and record reproduction status for DroidBot, GPTDroid, and MemoDroid. This scope leaves independent backbones, more commercial version pairs, and iOS/WebView subjects to future work. All counted objectives are read-only. The review package was prepared to exclude credentials, APKs, private dumps, and secrets; public release still requires a privacy and redistribution audit. The commercial health-companion subject remains anonymized because public naming permission has not been established.

\textbf{Alternative explanations and comparison limits.}
No experiment establishes that the specified lifecycle is preferable to simpler trace maintenance or strong retrieval with current-screen checks. The packaged memory implementation differs from the specification, so its enforcement and empirical benefit remain unverified. Adapted LLMDroid runs provide context, not a matched superiority test. Internal comparisons have slice-specific confounds; grounded SA shows that memory is not necessary for the recorded one-shot outcomes. Cross-backbone tests, unchanged-locator controls, and false-withdrawal measurements are needed before broader reliability claims.

\section{Related Work}

Android GUI testing has long studied random, model-based, search-based, state-restoration, reinforcement-learning, and platform-style exploration~\cite{monkey,droidbot,sapienz,stoat,combodroid,timemachine,ares,pan2020curiosity,droidmate2,ape}. LLM-assisted testers add semantic interaction, functional checking, coverage guidance, and dynamic memory~\cite{qtypist,gptdroid,auitestagent,llmdroid,lelante,memodroid,ncfllmoracle,trident}. \tool{} complements this line by asking how remembered GUI testing evidence is scoped, reused, audited, and retired across repeated sessions rather than ranking as a general exploration baseline.

General GUI agents, benchmarks, grounding systems, and memory-augmented agents target task completion, perception, or long-horizon reuse~\cite{appagent,mobileagentv2,mobileagente,agents,appagentx,aitw,androidworld,androidlab,guiodyssey,androiddaily,seeclick,cogagent,osatlas,uitars,omniparser,chainmemory,continuousmemory,raggui,echotrailgui,darwinianmemory,hymem,androtmem,executablememory,mementogui}.

\subsection{Positioning the Maintenance Question}
Random and model-based testing studies action generation and exploration. LLM-guided testing adds semantic interaction. Experience maintenance asks a complementary question: when an earlier interaction is reused, what makes it applicable now, and what evidence should revoke it? This distinction does not imply that earlier systems cannot implement such checks.

GPTDroid uses functionality-aware memory prompting to retain testing context and guide exploration~\cite{gptdroid}. LLMDroid combines autonomous exploration with LLM guidance when coverage growth slows~\cite{llmdroid}. MemoDroid organizes experience into episodic, reflective, and strategic memory and retrieves it into runtime prompts~\cite{memodroid}. These systems make memory itself an insufficient novelty claim. We focus on the applicability and retirement of remembered instructions after context or version changes, without claiming that these concerns are absent from all prior systems. Reproduction attempts and adapted executions are separated from internal controls in Section~\ref{sec:eval}. A flat-replay or maintenance-disabled variant of \tool{} is not a faithful implementation of MemoDroid or a substitute for a matched comparison. The study does not establish a state-of-the-art ranking.

Three memory layers, retrieval scoring, and role decomposition are not individually new ideas. Our contribution is an explicit testing-oriented maintenance contract, an audit of its implementation boundary, and selected task and update observations. The archive does not establish the benefit of the complete specified contract over simpler or model-managed policies.

\section{Conclusion}
We studied experience maintenance in repeated mobile GUI testing. The specified contract associates experience with scope and provenance, treats it as advisory, and requires a dedicated write path. The packaged implementation does not establish enforcement of the complete contract. Selected update cases illustrate stale-locator hazards, while the available summaries mix observed failures with policy-based prevention and offline retrieval. They do not establish a uniform end-to-end prevention rate or net savings. RQ2 is excluded from efficacy conclusions, and RQ4 route diagnostics do not establish general transfer acceleration. Priorities are implementing and testing the full contract, matched decision policies, independently checked outcomes, complete per-condition execution provenance, and broader update samples.

\section*{Use of Generative AI}
Generative-AI tools assisted drafting and language editing, and coding/debugging of analysis and artifact scripts. The package contains an explicitly identified Codex-B machine-side cross-check of selected annotation packets, not an independent human annotation or human ground truth. Later B-label revisions do not document an independent human protocol. The authors take responsibility for the text, results, references, and evidence provenance, including the validation limits disclosed in this paper.

\section*{Data Availability}
The evaluation records are retained in a local research archive. A public, stable, permission-cleared release URL is not yet established. The paper source contains no raw screenshots, device logs, or app binaries. Readers cannot currently independently rerun the full evaluation from this preprint alone. We distinguish archived observations from derived diagnostics and disclose unresolved validation limits in the corresponding result sections.

\begingroup
\footnotesize
\bibliographystyle{IEEEtran}
\bibliography{ref}
\endgroup

\end{document}